# IMPACT OF DIFFERENT SPREADING CODES USING FEC ON DWT BASED MC-CDMA SYSTEM

Saleh Masum[1], M. Hasnat Kabir[1], Md. Matiqul Islam[1], Rifat Ara Shams[1] and Shaikh Enayet Ullah[2]

[1]Department of Information and Communication Engineering
University of Rajshahi, Rajshahi-6205, Bangladesh.
[2]Department of Applied Physics, Electronic and Communication Engineering
University of Rajshahi, Rajshahi 6205, Bangladesh.

*ABSTRACT*

*The effect of different spreading codes in DWT based MC-CDMA wireless communication system is investigated. In this paper, we present the Bit Error Rate (BER) performance of different spreading codes (Walsh-Hadamard code, Orthogonal gold code and Golay complementary sequences) using Forward Error Correction (FEC) of the proposed system. The data is analyzed and is compared among different spreading codes in both coded and uncoded cases. It is found via computer simulation that the performance of the proposed coded system is much better than that of the uncoded system irrespective of the spreading codes and all the spreading codes show approximately similar nature for both coded and uncoded in all modulation schemes.*

## 1. INTRODUCTION

In recent years, Multi-Carrier Code Division Multiple Access (MC-CDMA), which has combined Orthogonal Frequency Division Multiplexing (OFDM) with Code Division Multiple Access (CDMA), has been received strong demand for future wireless communications [1]. MC-CDMA is a multiplexing scheme used in OFDM based telecommunication system which allows the system to support multiple users to access the wireless channel simultaneously by modulating and spreading their input data signals across the frequency domain using different spreading sequences [2]. MC-CDMA reaps the benefits of both OFDM and CDMA techniques which are the most important objectives in the 4[th] generation wireless mobile communication systems. For instance, it offers tremendous opportunity to accommodate many users within a given frequency band, provide high spectrum efficiency, improve the reliability and performance of wireless radio links, provide high speed data transmission rate, provide higher robustness on frequency selective interference, lower the symbol rate in each subcarrier so that a longer symbol duration makes it easier to quasi-synchronize the transmissions, provide narrow-band interference rejection etc. [1].

DOI : 10.5121/ijmnct.2012.2301                                                                                                                     1

However, nowadays Discrete Fourier Transform (DFT) based MC-CDMA is being replaced by spectrally efficient Discrete Wavelet Transform (DWT) based MC-CDMA because of its ability to provide higher bandwidth efficiency than the conventional MC-CDMA systems, reduce distortion in the reconstructed signal while retaining all the significant features present in the signal, provide better time and frequency localization, eliminate the need to append Cyclic Prefix (CP), improve Signal-to-Noise Ratio (SNR) performance, provide new dimensions for anti-fading and interference immunity etc. [3].

Our proposed paper is highlighted DWT based MC-CDMA system in different spreading codes with Forward Error Correction (FEC) over AWGN channel using different modulation schemes. There are various spreading codes which can be distinguished with respect to orthogonality and autocorrelation properties in order to optimally combine the multi-path signals of a particular user, cross-correlation properties for supporting multi-user communications, noise-like properties, implementation complexity and peak-to-average power ratio (PAPR) [4,5]. Here data symbols consisting of modulated bits are spread by spreading codes such as Walsh-Hadamard code, Orthogonal gold code and Golay code. The spreading code is a pseudorandom code that effectively represents each bit of an information signal by multiple RF information signals over a frequency band that is much wider than the information signal [6]. It is also used for dispreading and subsequent data recovery. Moreover FEC, which is accomplished by adding redundancy to the transmitted information using a predetermined algorithm, can prevent the miscommunication of a message by working to correct any mistakes or scrambling of the message that occurs during transmission without relying on retransmissions of data. A particular type of error control code, called the Golay code, is introduced in this paper which enhances the reliability of communication on a noisy data link [7]. The objectives of this paper are to identify the behavior of different spreading codes on DWT based MC-CDMA system and compare the results with coded proposed system.

The rest of this paper is organized as follows: Section 2 describes about the different spreading codes. Section 3 describes regarding FEC code (Golay code). The system model is described in section 4. In section 5, the simulation results and discussion of different spreading codes using Forward Error Correction on DWT based proposed MC-CDMA system is presented. Finally the conclusion part is added in section 6
.
## 2. SPREADING CODES

Spread spectrum is a means of transmission in which signal occupies a bandwidth much wider than the minimum bandwidth required to transmit the information being sent. The frequency spectrum of the original information signal is spread using user-specific signature sequences uncorrelated with that signal and the synchronized replica of the signature sequences is then correlated with the received signal by the receiver in order to recover the original information [4]. It has a number of advantages such as, low power spectral density, interference rejection, low probability of intercept, multiple access, privacy due to unknown random codes, reduction of multi-path effects, anti-jamming, diversity reception, high resolution ranging and accurate universal timing etc. [8]. The spreading techniques in MC-CDMA differ in the selection of the spreading code and the type of spreading. Spreading codes are one of the major elements within the whole MC-CDMA system which are combined with the data stream to be transmitted in such a way that the bandwidth required is increased, the signal is spread and the benefits of the spread spectrum system can be gained. Co-existing of multiple coded channels at the same time on the same frequency can be possible using different spreading codes for each information signal. In our proposed DWT based MC-CDMA system, Walsh-Hadamard code, Orthogonal gold code and Golay complementary sequences are used as the spreading codes. A brief description is presented here.

## 2.1. Walsh-Hadamard Code

The name of this code comes from the American mathematician Joseph Leonard Walsh and the French mathematician Jacques Hadamard. It is a decodable code which recovers parts of the original message with high probability, while only looking at a small fraction of the received word. Orthogonality is the most important property of Walsh–Hadamard codes which provides the zero cross-correlation between any two Walsh–Hadamard codes of the same set when the system is synchronized. It is a linear code over a binary alphabet that maps messages of length n to codewords of length $2^n$. Each non-zero codeword has Hamming weight of exactly $2^{n-1}$, which implies that the distance of the code is also $2^{n-1}$. In standard coding theory notation, this means that the Walsh–Hadamard code is a $[2^n, n, 2n/2]_2$-code [9-10]. The generating algorithm is as follows.

$$H_{2N} = \begin{bmatrix} H_N & H_N \\ H_N & \overline{H_N} \end{bmatrix}$$

Where N is a power of 2 and N can be defined from the following recurrent rule [8]:

$$H_0 = [1];$$

$$H_1 = \begin{bmatrix} Ho & Ho \\ Ho & -Ho \end{bmatrix};$$

$$H_{i+1} = \begin{bmatrix} Hi & Hi \\ Hi & -Hi \end{bmatrix}; \quad i = 1 \ldots \log_2(N) - 1;$$

## 2.2. Orthogonal Gold Code

In telecommunication and satellite navigation, a spatial type of code called Gold code is used to provide bounded small cross-correlations within a set, which is useful when multiple devices are broadcasted in the same range [11]. The name comes from its inventor Robart Gold. Gold code is used in MC-CDMA as chipping sequences that allows several callers to use the same frequency, resulting in less interference and better utilization of the available bandwidth. A set of n Gold sequences can be generated from a preferred pair of maximum length m-sequences of the same length $2^n - 1$, such that their absolute cross-correlation is less than or equal to $2^{(n+2)/2}$, by modulo-2 addition of the first preferred m-sequence with the n cyclically shifted versions of the second preferred m-sequence. Gold code has a three-valued autocorrelation and cross-correlation function with values {-1, -t(m), t(m) - 2}, where

$$t(m) = \begin{cases} 2^{(m+1)/2} + 1 & \text{for m odd} \\ 2^{(m+2)/2} + 1 & \text{for m even} \end{cases}$$

One can find that many cross-correlation values of Gold code is "-1" for many code offsets. This suggests that it may be possible to make cross-correlation values to "0" by padding an additional "0" to the original Gold code and $2^n$ orthogonal codes can be obtained by this simple zero padding.

### 2.3. Golay Complementary Sequences

The complementary sequences (CS) was introduced by Marcel J. E. Golay in 1949 in the context of infrared spectrometry, are pairs of sequences with the useful property that the sum of their out-of-phase aperiodic autocorrelation coefficients vanishes at all delays other than zero [12]. Let ($a_0$, $a_1$, ..., $a_{N-1}$) and ($b_0$, $b_1$, ..., $b_{N-1}$) be a pair of bipolar sequences, meaning that $a(k)$ and $b(k)$ have values +1 or −1. Let the aperiodic autocorrelation function of the sequence **x** be defined by,

$$R_x(k) = \sum_{j=0}^{N-k-1} x_j x_{j+k}, \qquad 0 \leq k \leq N-1.$$

Then the pair (a, b) is called a Golay Complementary Pair (GCP) if:

$R_a(k) + R_b(k) = 0, \qquad$ for $k = 1, ..., N-1$.

Each member of a GCP is known as a Golay Complementary Sequence (GCS). The perfect autocorrelation property of Golay complementary sequences has proved to be of value in a variety of applications including ultrasound imaging, coded aperture spectroscopy and anti-aliasing in the fields of physics, combinatory and telecommunications [13].

## 3. FORWARD ERROR CORRECTION USING GOLAY CODE

Forward Error Correction (FEC) is a type of digital signal processing that improves data reliability by introducing a redundant data known as Error-Correcting Code (ECC) into a data sequence prior to transmission or storage. This structure enables the receiver to detect errors caused by corruption that may occur anywhere in the message in the channel as well as the receiver. It corrects these errors without requesting retransmission of the original information [14]. There are two main types of FEC codes in coding theory: block codes and convolution codes. A block code is a code which does not contain any additional information, works on fixed-size blocks of bits or symbols of predetermined size and corrects errors that occurs in the transmission of the code. The encoder for a block code is memoryless that is the digits in each codeword is independent of any information contained in previous codewords. However, practical block codes are decoded in polynomial time to their block length. On the other hand, convolution code is a code that works on bit or symbol streams of arbitrary length where the digits of a codeword depend on the codewords that are encoded previously during a fixed length of time while viterbi algorithm is used for decoding [7]. In our proposed system we have used binary Golay code that encodes 12 bits into 23 bits, denoted by (23, 12) as error correcting code.

It is a type of linear binary block code used in real-time applications, mathematics and electronics engineering that requires low latency and short codeword length. According to Golay code

$$\binom{23}{0} + \binom{23}{1} + \binom{23}{2} + \binom{23}{3} = 2^{11} = 2^{23-12}$$

which indicates the possible existence of a (23, 12) perfect binary code that is capable of correcting any combination of three or fewer random errors in a block of 23 elements [7]. This (23, 12) Golay code can be generated either by

$g_1(X) = 1 + X^2 + X^4 + X^5 + X^6 + X^{10} + X^{11}$

or $\quad g_2(X) = 1 + X + X^5 + X^6 + X^7 + X^9 + X^{11}$

Both polynomials $g_1(X)$ and $g_2(X)$ are factors of $X^{23} + 1$. Where,

$X^{23} + 1 = (1+X)\, g_1(X)\, g_2(X)$

A codeword structure of the binary Golay code is formed by taking 12 information bits and appending 11 check bits which is derived from a modulo-2 division, as with the Cyclic Redundancy Check (CRC).

| Golay [23,12] Codeword | |
|---|---|
| Check bits | Information bits |
| XXX XXXX XXXX | XXXX XXXX XXXX |

The interesting property of Golay codeword is cyclic invariance which means that if we take a 23-bit Golay codeword and cyclically shifted it by any number of bits, the result is also a valid Golay codeword. In other word, if we take a 23-bit Golay codeword and invert it, the result is also a valid Golay codeword.

## 4. SYSTEM MODEL

The main objective of this work is to simulate the DWT based MC-CDMA system by utilizing different spreading codes with FEC. The block diagram of the entire system is depicted in figure 1 where the designated block represents their specific operation.

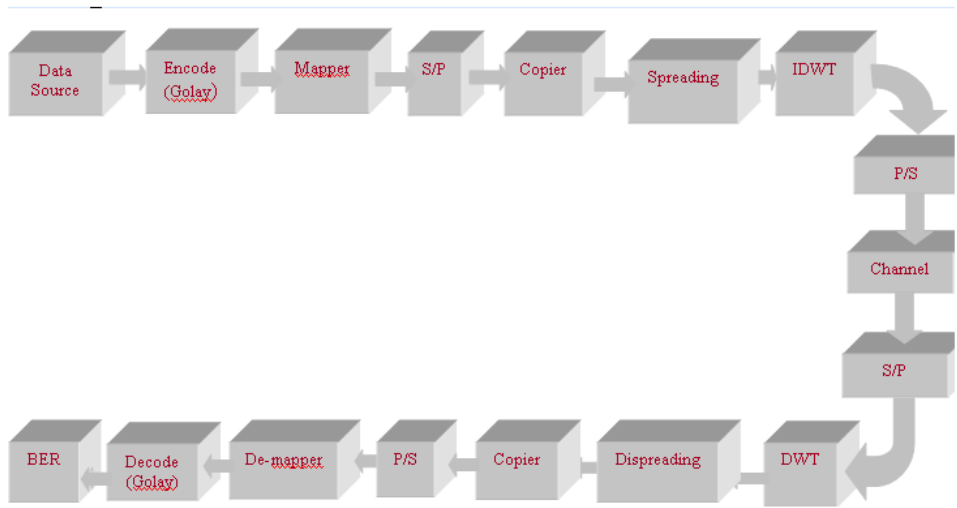

Figure 1. Simulation block diagram of proposed DWT based MC-CDMA system using different spreading codes with implementation of FEC.

At first the binary bit stream is synthetically generated using the standard random bit generator which is encoded using cyclic Golay code in order to detect and remove errors from received data. The encoded output is then converted into complex symbols using BPSK, QPSK, DBPSK and DQPSK digital modulation. These symbols are converted into serial to parallel form and are fed into copier section where they are multiplied with assigned individual Walsh-Hadamard code, Orthogonal Gold code and Golay complementary sequences code to spread the signal and to increase the required bandwidth. Then wavelets are generated and signal synthesis is performed using inverse wavelet transformation. Parallel to serial conversion is performed again to make the signals ready for transmission and the transmitted signals are passed through the AWGN channel. At the receiver side, inverse operations are performed to decode the received sequences of information bits to retrieve the original data source signal. The probability of Bit Error Rate (BER) of the system as a function of Signal-to-Noise Ratio (SNR) is calculated by comparing the transmitted and the received data.

## 5. SIMULATION RESULTS AND DISCUSSION

In our proposed DWT based MC-CDMA system, the comparisons are based on simulation which has been done using MATLAB-7.5. The parameters of the simulation model are summarized in Table 1. In different spreading codes, that is Walsh-Hadamard code, Orthogonal Gold code and Golay complementary sequences, the bit error probability performance of coded system is compared with respective uncoded system in the presence of Additive White Gaussian Noise (AWGN) using different modulation schemes.

Table 1: The parameters of the simulation model

| Parameter | Parameter values |
|---|---|
| User | 7 |
| SNR range | 0-20 dB |
| Wavelets | Haar, Daubechies, Bi-orthogonal |
| Vanishing order of Daubechies | 2 |
| Vanishing order of Bi-orthogonal | 2 |
| DWT and IDWT size | 256 |
| Digital modulation | BPSK, QPSK, DBPSK, DQPSK |
| Wireless channel | AWGN |
| Forward Error Correction (FEC) code | Golay code (23, 12) |
| Spreading code | Walsh-Hadamard code, Orthogonal Gold code, Golay complementary sequence |
| Spreading factor of spreading codes | 8 and 16 |

The BER performance of Golay coded and uncoded DWT based MC-CDMA system in different spreading codes over AWGN channel using BPSK, DBPSK, QPSK and DQPSK modulation schemes is presented by Fig. 2, 3, 4 and 5, respectively for wide range of SNR from 0dB to 20dB.

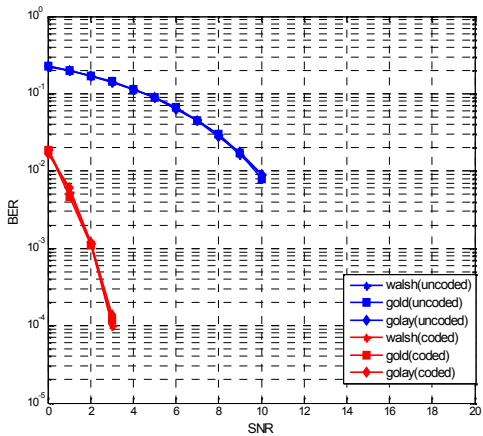

Figure 2. Comparison of coded and uncoded DWT based MC-CDMA for Different spreading codes using BPSK modulation over AWGN channel.

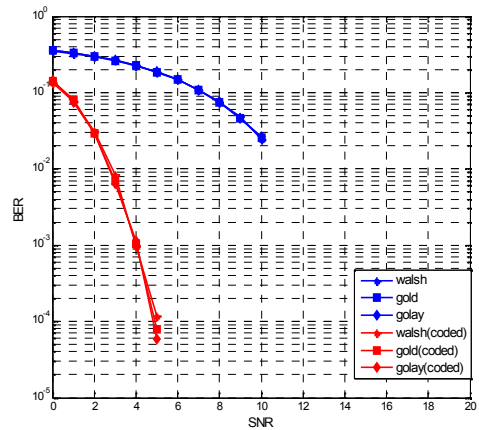

Figure 3. Comparison of coded and uncoded DWT based MC-CDMA for Different spreading codes using DBPSK modulation over AWGN channel.

Figure 2 illustrates the comparison of the performance of different spreading codes using BPSK modulation scheme over AWGN channel with implementation of FEC. In terms of performance, there are no major differences can be observed between Walsh-Hadamard code, Orthogonal Gold code and Golay complementary sequences for both coded and uncoded system. However, in comparison between coded and uncoded system, much better performance is found for coded system as it is expected. Nevertheless, 3 orders of magnitude better performance is achieved by coded system in contrast to uncoded one at 3 SNR.

The simulation result in figure 3 shows that the BER is inversely proportional to the SNR as expected. Here the coded DWT based MC-CDMA using DBPSK modulation over AWGN channel has better BER performance than uncoded system irrespective of spreading codes. There are 2 orders of magnitude better performance is obtained by coded system. On the other hand, careful observation between three spreading codes under coded system indicate that the Golay spreading has a small impact in BER performance which is shown lower value. The performance of different spreading codes in DBPSK is shown similar nature as it is found in BPSK in both coded and uncoded cases.

In comparison between figure 2 and figure 3, it is found that the BPSK has better performance than that of DBPSK in both coded and uncoded cases. This conclusion supports the theory of modulation scheme. In DBPSK, the output symbol phase corresponds to the phase difference between the present and the previous symbols. As a result, the symbol noise is doubled at DBPSK in contrast to the phase noise of a single symbol as used in coherent modulation like BPSK.

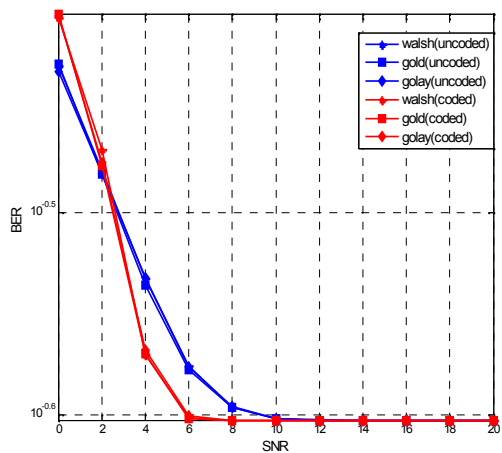 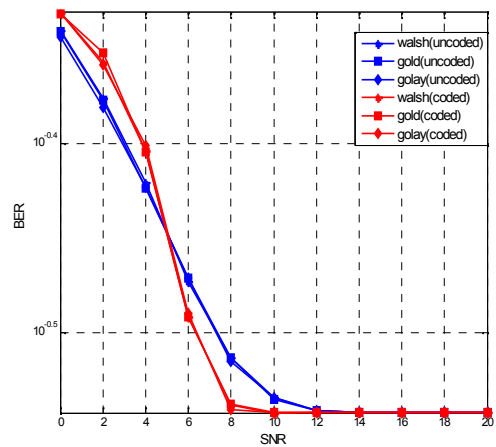

Figure 4. Comparison of coded and uncoded DWT based MC-CDMA for Different spreading codes using QPSK modulation over AWGN channel.

Figure 5. Comparison of coded and uncoded DWT based MC-CDMA for Different spreading codes using DQPSK modulation over AWGN channel

It is observed from figure 4 and figure 5 that the performance in both modulations is similar nature as usual. Here an interesting point can be noted that the coded performance is degraded initially up to nearly SNR 2 for QPSK and SNR 5 for DQPSK, respectively but later the performance is increased slightly up to SNR 10 for QPSK and SNR 12 for DQPSK, respectively. A minor difference in terms of performances is shown between three used spreading crowds in both coded and uncoded cases. The overall performance of the system for QPSK modulation scheme is slightly better in contrast to the performance for DQPSK modulation scheme.

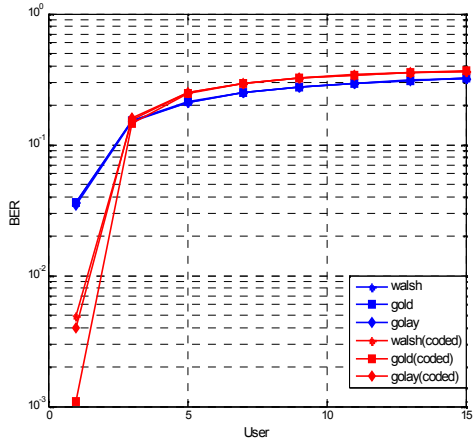

Figure 6. Comparison of coded and uncoded DWT based MC-CDMA in different spreading codes for different users using BPSK modulation over AWGN channel where SNR is -10dB.

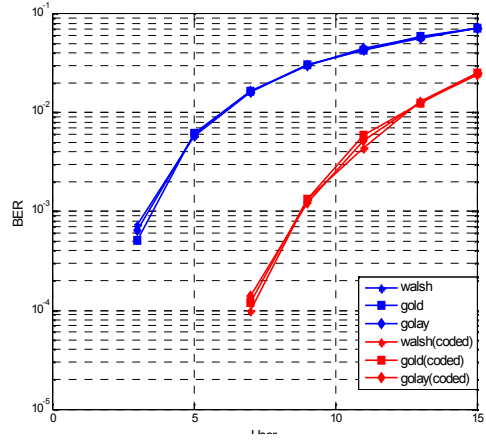

Figure 7. Comparison of coded and uncoded DWT based MC-CDMA in different spreading codes for different users using BPSK modulation over AWGN channel where SNR is 0dB.

In AWGN channel we consider another scenario where the BER performance is measured against single SNR level for different users. From figure 6 and figure 7, it is seen that if the number is increased, BER is also increased. In other word, BER is directly proportional to the number of users. It can be pointed out here that the number of users has a great impact on the performance. The performance is better in low number of users. However, there is another point should be mentioned here that the coded system shows better performance than that of uncoded one.

## 6. CONCLUSION

An approach is presented in this paper that applies different spreading codes along with FEC in DWT based MC-CDMA system using different modulation schemes over AWGN channel. From simulation results it is seen that the spreading codes (Walsh-Hadamard code, Orthogonal Gold code and Golay complementary sequence) bear almost similar performances in AWGN channel. The strong impact of spreading codes is not observed for the transmission of information in the proposed method. However, the encoded proposed system is better than that of uncoded system over AWGN channel irrespective of modulation schemes and spreading codes. It is observed that the performance of BPSK modulation technique is the best among others while the values of SNR are varied for constant users. On the other hand, SNR level of 0dB has quite better performance than SNR level of -10dB while the number of users are varied.